\documentclass[final,5p,times,twocolumn]{elsarticle}
\usepackage{graphicx}
\usepackage{dcolumn}
\usepackage{bm}
\usepackage{amsmath}
\usepackage{amssymb}
\usepackage{latexsym}
\usepackage{epsfig}
\usepackage{amsbsy}
\usepackage{array}
\usepackage{amssymb}
\usepackage{setspace}
\usepackage{bm}
\usepackage{epstopdf}
\usepackage{subfigure}
\biboptions{sort&compress}

\journal{Journal of \LaTeX\ Templates}

\begin{document}

\begin{frontmatter}
\title{Magnon squeezing enhanced entanglement in a cavity magnomechanical
system}
\author[mymainaddress]{Ming-Song Ding}
\author[mymainaddress]{Ying Shi}
\author[mymainaddress]{Yu-jie Liu}
\author[mymainaddress]{Li Zheng\corref{mycorrespondingauthor}}
\cortext[mycorrespondingauthor]{Corresponding author}
\ead{dingms@dlpu.edu.cn}

\address[mymainaddress]{Basic Education Department, Dalian polytechnic University, Dalian
116034, China}

\begin{abstract}
We investigate the generation of the entanglement in a cavity
magnomechanical system, which consists of three modes: a magnon mode, a
microwave cavity mode and a mechanical vibration mode, the couplings of the
magnon-photon and the magnon-phonon are achieved by the magnetic dipole
interaction and the magnetostrictive interaction, respectively. By
introducing a squeezing of the magnon mode, the magnon-photon and the
magnon-phonon entanglements are significantly enhanced compared with the
case without inserting the magnon squeezing. We find that an optimal
parameter of the squeezing exists, which yields the maximum entanglement.
This study provides a new idea for exploring the properties of quantum
entanglement in the the cavity magnomechanical systems, and may have some
potential applications in the quantum state engineering.

\end{abstract}

\end{frontmatter}

\section{Introduction}


Cavity magnomechanical system (CMM system) is becoming a new hotspot arising
in recent years, this system consists of a microwave cavity and a
yttrium-iron-garnet (YIG) sphere. Because of the high spin density and the
strong spin-spin exchange interaction, the magnons which are embodied by a
collective motion of the uniform of spins can be strongly couple to cavity
microwave photons in the YIG sphere\cite{lachance2019hybrid,k2}, and the
coupling strength between them can even achieve the ultrastrong coupling 
\cite{k201}. Simultaneously, the magnon mode can also couple to a mechanical
vibration mode with the magnetostrictive interaction (the radiation pressure
like in the optomechanical systems), which realizes photon-phonon coupling.
Comparing with the optomechanical systems \cite%
{k4,k5,li2020noise,zeng2022suppressing,zhang2021measurement}, the CMM
systems have the advantages of high adjustability and low loss. Hence this
system can provide a promising vision for the realization of quantum
information theory and macroscopic quantum states in hybrid quantum systems 
\cite%
{k61,k7,k62,k8,k9,k21,li2021magnetic,k11,k22,k222,k223,k226,cheng2022generation,k225,yuan2020enhancement,k701}%
.

As a fundamental physical resource of the quantum information processing,
the quantum entangled states are widely used in the quantum computing,
quantum information processing and others, the study of entanglement emerges
in endlessly in the mechanical systems. At the same time, people are also
thinking about how to enhance this important physical resource\cite%
{hofer2015entanglement,wang2014nonlinear,k37,hu2020entanglement,li2019dynamics}%
. On the other hand, it is a prerequisite for observing various quantum
effects in a mechanical system to make the quantum system cold to close to
its ground state, and it is helpful for the generation of the entanglement 
\cite{martin2004ground,k224,lai2021domino}. Li et al. have done a novel work 
\cite{asjad2022magnon} and they found that the magnon squeezing can enhanced
the ground-state cooling in the CMM system. And the squeezing of the magnon
mode can also use to enhance the nonlinearity and entanglement.

Here, We utilize a CMM system with the magnon squeezing to study the
entanglement inside. The magnetostrictive interaction and magnetic dipole
interaction mediate the magnon-phonon coupling and the photon-magnon
coupling, respectively. We provide a protocol to enhance the continuous
variable entanglement between the magnon mode and cavity mode (or the phonon
mode) via the squeezing of the magnon mode, with the squeezing parameter $%
\chi $ and the phase $\theta $. We show the entanglement can be
significantly enhanced, and the optimal values of the phase and the
squeezing parameter corresponding to the maximum entangleme are given.
Furthermore, the effect of temperature on entanglement is also discussed.

The paper is organized as follows. In Sec. II, we show the Hamiltonian and
dynamical equations of the whole system. In Sec. III, we compare the
entanglement in the cases with and without inserting the magnon squeezing.
It shows that remarkable improvement of the entanglement can be achieved
with the enhancement of nonlinear effect caused by the squeezing. Finally, a
concluding summary is given in Sec. IV.

\section{\protect\bigskip Theoretical model and dynamical equation}

As shown in Fig. 1, we consider a three-mode CMM system, which consists of a
magnon mode (the uniform-precession Kittel mode in a YIG sphere \cite{k61}),
a microwave cavity mode, and a mechanical vibration mode. When the frequency
of the Kittel-mode magnons resonates with the frequency of the microwave
photons, the magnon mode and optical mode can be strongly coupled.
Simultaneously, the magnon mode can couple to the mechanical vibration mode
by the magnetostrictive interaction. Because\ the size of the YIG sphere
(the diameter of the YIG sphere is generally $10^{2}\mu $m to $1$mm) is much
smaller than the wavelength of the microwave field, the photon-phonon
coupling caused by the radiation pressure is ignored. Besides, an external
microwave driving field is introduced to effectively enhance the
magnon-phonon coupling \cite{k62,k7}. With a rotating frame at the frequency
of the optical driving field $\omega _{d}$, the total Hamiltonian of this
system reads ($\hbar =1$)

\begin{eqnarray}
H &=&\Delta _{a}a^{\dagger }a+\Delta _{m}m^{\dagger }m+g_{ma}(a^{\dagger
}m+m^{\dagger }a)  \notag \\
&&+\frac{\omega _{b}}{2}(x^{2}+p^{2})+g_{mb}m^{\dagger }mx  \label{eq01} \\
&&+i\varepsilon _{d}(m^{\dagger }e^{-i\omega _{d}t}-me^{i\omega _{d}t}) 
\notag \\
&&+i\chi (m^{\dagger 2}e^{i\theta }-m^{2}e^{-i\theta }),
\end{eqnarray}%
where $a(a^{\dagger })$ and $m(m^{\dagger })$ are the annihilation(creation)
operators of the cavity mode and the uniform magnon mode at the frequency $%
\omega _{a}$ and $\omega _{m}$, respectively ($[O,O^{\dagger }]=1,O=a,m$). $%
\Delta _{a(m)}=\omega _{a(m)}-\omega _{d}$ represents the detuning of
frequencies $\omega _{a(m)}$ and $\omega _{d}$. The magnon frequency $\omega
_{m}$ can be easily adjusted by altering the external bias magnetic field $H$
via $\omega _{m}=\gamma _{g}H$, with the the gyromagnetic ratio $\gamma
_{g}=28$GHz/T. $x$ and $p$ are the dimensionless position and momentum
quadrature of the mechanical vibration mode with the frequency $\omega _{b}$
($[x,p]=i$). The coupling rate of the magnon-cavity interaction and
magnon-phonon interaction are given by $g_{ma}$ and $g_{mb}$, respectively.
Under the assumption of the low-lying excitations, the Rabi frequency $%
\varepsilon _{d}=\frac{\sqrt{5}}{4}\gamma _{g}\sqrt{N_{t}}B_{0}$ indicates
the coupling rate of the magnon mode and the driving field with the
amplitude $B_{0}$, with the total number of the spins $N_{t}$ $=\rho V$ ($V$
is the volume of the YIG sphere) and the spin density $\rho =$ $4.22\times
10^{27}$m$^{-3}$. It is worth noting that the last term in the total
Hamiltonian is the squeezing of the magnon mode, with the squeezing
parameter $\chi $ and the phase $\theta $ \cite{sohail2022enhanced}. So far,
there are several methods to achieve the magnon squeezing \cite{33,21,47}.

\begin{figure}[tbp]
\centering\includegraphics[width=8cm]{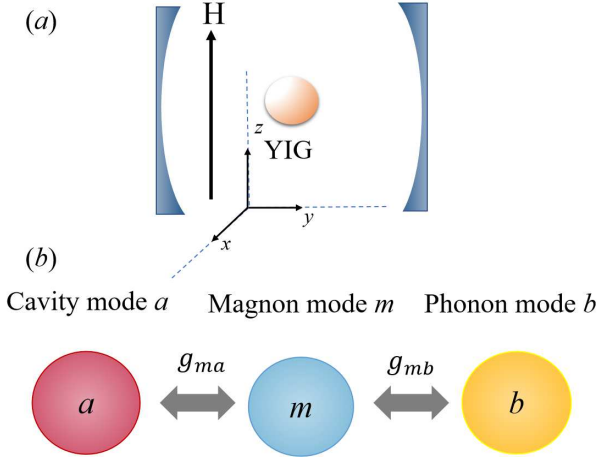} 
\caption{(a) Schematic illustration of the system, a YIG sphere is placed in
a microwave cavity. A bias magnetic field $H$ along the $z$-axis is used to
realize magnon-photon coupling by magnetic dipole interaction. Owing to
magnetostrictive effect, the vibrating motion of the surface of the YIG
sphere is considered as a mechanical resonator. (b) The equivalent
mode-coupling model.}
\end{figure}

By including the dissipation and input noise of each mode, the\ quantum
Langevin equations (QLEs) are given by

\begin{eqnarray}
\dot{a} &=&(-i\Delta _{a}+\Gamma )a-ig_{ma}m+\sqrt{2\kappa _{a}}a^{in}, 
\notag \\
\dot{m} &=&(-i\Delta _{m}-\kappa _{m})m-ig_{ma}a+\varepsilon _{d}  \notag \\
&&-ig_{mb}mx+2\chi m^{\dagger }e^{i\theta }+\sqrt{2\kappa _{m}}m^{in}, 
\notag \\
\dot{x} &=&\omega _{b}p,  \label{eq02} \\
\dot{p} &=&-\omega _{b}x-\gamma _{b}p-g_{mb}m^{\dagger }m+\xi ,  \notag
\end{eqnarray}%
where $\kappa _{m}$ and $\gamma _{b}$ represent the dissipation rates of
magnon mode and mechanical mode, respectively. As the aim is to enhance
entanglement, an optical effective gain $\Gamma =-\kappa _{a}+G_{a}$ (active
optical cavity) is considering in this CMM system \cite{ding2021enhanced},
where $G_{a}$ is the gain of cavity mode and it is greater than the
dissipation of the cavity field $\kappa _{a}$. This optical gain can be
achieved by in many methods \cite{42,43}. $a^{in},m^{in}$ and $\xi $ are
input noise operators of cavity, magnon and mechanical modes, with a
Markovian approximation, the noise operators satisfy: $\left\langle
a^{in}(t)a^{in\dagger }(t^{\prime })\right\rangle =(n_{a}+1)\delta
(t-t^{\prime })$, $\left\langle a^{in\dagger }(t)a^{in}(t^{\prime
})\right\rangle =n_{a}\delta (t-t^{\prime })$, $\left\langle
m^{in}(t)m^{in\dagger }(t^{\prime })\right\rangle =(n_{m}+1)\delta
(t-t^{\prime })$, $\left\langle m^{in\dagger }(t)m^{in}(t^{\prime
})\right\rangle =n_{m}\delta (t-t^{\prime })$, $\left\langle \xi (t)\xi
^{\dagger }(t^{\prime })\right\rangle =(n_{b}+1)\delta (t-t^{\prime })$ and $%
\left\langle \xi ^{\dagger }(t)\xi (t^{\prime })\right\rangle =n_{b}\delta
(t-t^{\prime })$, with $n_{\mu }=(e^{\hbar \omega _{\mu }/k_{B}T}-1)$($\mu
=a,m,b$), $k_{B}$ is the Boltzmann constant and $T$ the environmental
temperature.

\section{Entanglement generation}

According to the strong driving field, each Heisenberg operator can
rewritten as a sum of the steady-state mean value and the corresponding
quantum fluctuation, i.e., $O(t)=O_{s}+\delta O(t)$ $(O=a,m,x,p)$, then we
can linearize this system. The linearized QLEs can be written in a compact
form $\dot{u}(t)=Au(t)+n(t)$, where quadratures of the quantum fluctuations
is rewritten as $u(t)=[\delta X_{1}$, $\delta X_{2}$, $\delta Y_{1}$, $%
\delta Y_{2}$, $\delta x$, $\delta p]^{T}$, and we define $\delta
X_{1}=(\delta a+\delta a^{\dagger })/\sqrt{2}$, $\delta X_{2}=i(\delta
a^{\dagger }-\delta a)/\sqrt{2}$, $\delta Y_{1}=(\delta m+\delta m^{\dagger
})/\sqrt{2}$and $\delta Y_{2}=i(\delta m^{\dagger }-\delta m)/\sqrt{2}$.
Defining the vector of noise $n(t)=[\sqrt{2\kappa _{a}}X_{1}^{in}(t)$, $%
\sqrt{2\kappa _{a}}X_{2}^{in}(t)$, $\sqrt{2\kappa _{m}}Y_{1}^{in}(t)$, $%
\sqrt{2\kappa _{m}}Y_{2}^{in}(t)$, $0$, $\xi (t)]^{T}$ and the correlation
matrix is given by

\begin{equation}
A=\left( 
\begin{array}{cccccc}
\Gamma & \Delta _{a} & 0 & g_{ma} & 0 & 0 \\ 
-\Delta _{a} & \Gamma & -g_{ma} & 0 & 0 & 0 \\ 
0 & g_{ma} & \mu_{+} & \nu_{+} & -G & 0 \\ 
-g_{ma} & 0 & \nu_{-} & \mu_{-} & 0 & 0 \\ 
0 & 0 & 0 & 0 & 0 & \omega _{b} \\ 
0 & 0 & 0 & G & -\omega _{b} & -\gamma _{b}%
\end{array}%
\right) ,  \label{eq10}
\end{equation}%
where $G=i\sqrt{2}g_{mb}m_{s}$ is the coherent-driving-enhanced
magnomechanical coupling strength, $\mu _{\pm }=-\kappa _{m}\pm 2\chi \cos
\theta $, $\nu _{\pm }=\pm \tilde{\Delta}_{m}+2\chi \sin \theta $ and $m_{s}$
is

$\ \ \ \ \ \ \ \ \ \ \ \ \ \ \ \ \ \ \ \ \ $%
\begin{equation}
m_{s}=\frac{\varepsilon _{d}(i\Delta _{a}-\Gamma )}{g_{ma}^{2}+(i\Delta
_{a}-\Gamma )\beta },  \label{eq09}
\end{equation}%
$\bigskip $with the effective magnon-drive detuning $\tilde{\Delta}%
_{m}=\Delta _{m}+g_{mb}q_{s}$ and $\beta =(i\tilde{\Delta}_{m}-2\chi \sin
\theta )+(\kappa _{m}-2\chi \cos \theta )$.

For this system, due to the linearized dynamics of the system and the
Gaussian input noise operators, the entanglement of magnon-cavity and
magnon-phonon can be measure by the continuous variable (CV) entanglement,
we choose a standard ensemble method (logarithmic negativity $E_{\mathcal{N}%
})$ \cite{k43}. When the stability conditions are satisfied, one gets the
following equation for the steady-state correlation matrix (CM) \cite{k38}:

\begin{equation}
AV+VA^{T}=-D,  \label{eq11}
\end{equation}%
with the diffusion matrix $D=$diag$[\kappa _{a}(2n_{a}+1)$, $\kappa
_{a}(2n_{a}+1)$, $\kappa _{m}(2n_{m}+1)$, $\kappa _{m}(2n_{m}+1)$, $0$, $%
\gamma _{b}(2n_{b}+1)]$. In the CV case, the logarithmic negativity is
defined as \cite{djorwe2014robustness}

\begin{equation}
E_{\mathcal{N}}=\max [0,-\ln 2\eta ],  \label{eq13}
\end{equation}%
where $\eta =2^{-1/2}\{\Sigma (V)-[\Sigma (V)^{2}-4\det V_{s}]^{1/2}\}^{1/2}$%
, with $\Sigma (V)=\det A+\det B-2\det C$. The matrix elements of the
reduced $4\times 4$ submatrix $V_{s}$ depend on the one of the bipartite
entanglements: $E_{\mathcal{N}\text{,}am}$ and $E_{\mathcal{N}\text{,}bm}$
denote the photon-magnon and magnon-phonon entanglement, respectively. And $%
V_{s}$ can be written in the following form 
\begin{equation}
V_{s}=\left( 
\begin{array}{cc}
A & C \\ 
C^{T} & B%
\end{array}%
\right) .  \label{eq14}
\end{equation}

\begin{figure}[tbp]
\centering\includegraphics[width=8.5cm]{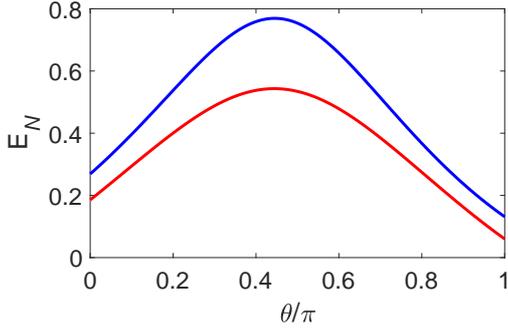} 
\caption{The entanglements $E_{\mathcal{N}\text{,}am}$ and $E_{\mathcal{N}%
\text{,}bm}$ as functions of the phase $\protect\theta $. The red line
denotes $E_{\mathcal{N}\text{,}am}$ (cavity-magnon entanglement) and the
blue line denotes $E_{\mathcal{N}\text{,}bm}$ (phonon-magnon entanglement).
We set $\protect\chi /2\protect\pi =0.4$MHz, $g_{ma}/2\protect\pi =3.5$MHz,
red line($g_{ma}/2\protect\pi =3.5$MHz), blue line($g_{ma}/2\protect\pi =4.7$%
MHz) and the temperature $T=12$mK. See text for the other parameters.}
\end{figure}

\begin{figure}[tbp]
\centering\includegraphics[width=8.5cm]{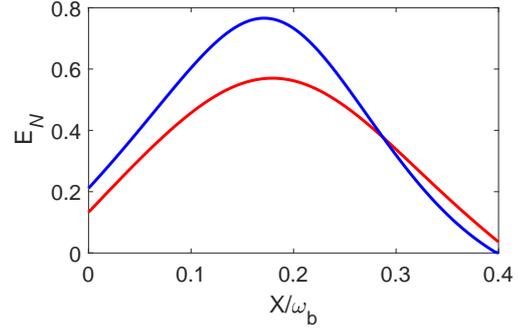} 
\caption{The entanglements $E_{\mathcal{N}\text{,}am}$ and $E_{\mathcal{N}%
\text{,}bm}$ as functions of the squeezing parameter $\protect\chi $. The
red line denotes $E_{\mathcal{N}\text{,}am}$ (cavity-magnon entanglement)
and the blue line denotes $E_{\mathcal{N}\text{,}bm}$ (phonon-magnon
entanglement). We set $\protect\theta =0.8\protect\pi $, red line($g_{ma}/2%
\protect\pi =3.5$MHz), blue line($g_{ma}/2\protect\pi =4.7$MHz) and the
other parameters we chose are the same as those in Fig.2.}
\end{figure}

One can now characterize the entanglement through Eq. (\ref{eq13}). In Fig.
2 we plot the bipartite entanglements $E_{\mathcal{N}\text{,}am}$ and $E_{%
\mathcal{N}\text{,}bm}$ versus the phase $\theta $, it can be seen that the
optimal phase for the entanglement is $\theta \simeq 0.44\pi $, the
entanglement in $\theta \in \lbrack \pi ,2\pi ]$ is much weak, which is not
discussed here. Then the following experimentally achievable parameters in
this work are given \cite{k61,k62}: $\Delta _{a}/2\pi =\tilde{\Delta}%
_{m}/2\pi =10$MHz (blue detuning), $\omega _{b}/2\pi =10$MHz, $\gamma
_{b}/2\pi =10$Hz, $\kappa _{m}/2\pi =1$MHz, $\kappa _{a}/2\pi =0.5$MHz, $%
g_{mb}/2\pi =0.2$Hz and $\varepsilon _{d}\simeq 3.5\times 10^{14}$Hz,
corresponding to the drive power $P=0.84$mW. Fig. 3 shows that the squeezing
of the magnon mode is helpful to enhance the bipartite entanglements.
Compared with the case of without the squeezing ($\chi =0$), the
entanglement $E_{\mathcal{N}\text{,}am}$ and $E_{\mathcal{N}\text{,}bm}$ can
be significantly enhanced, and it can be seen that the maximum of the
entanglement with the squeezing increases by 320$\%$ for $E_{\mathcal{N}%
\text{,}am}$ and 260$\%$ for $E_{\mathcal{N}\text{,}bm}$. However, when the
value of $\chi $ continues to increase and exceed the optimal value, the
entanglement will decrease. Physically, the gradual increase of $\chi $ from 
$0$ means that the nonlinearity of the system is enhanced, resulting in the
increase of entanglement. However, when $\chi $ continues to increase and
exceed the optimal value, the noise of magnons becomes a significant
effective thermal bath for the mechanical mode, and its negative impact on
the generation of entanglement exceeds the positive impact of the squeezing.

\begin{figure}[tbp]
\centering\includegraphics[width=6cm]{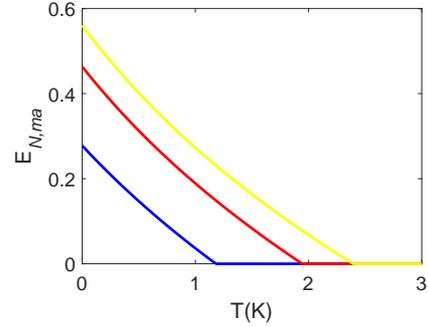} 
\caption{The entanglements $E_{\mathcal{N}\text{,}am}$ as functions of the
temperature $T$. The yellow line denotes the squeezing parameter $\protect%
\chi =0.15\protect\omega _{b}$, the red line denotes $\protect\chi =0.1%
\protect\omega _{b}$ and the blue line denotes $\protect\chi =0.04\protect%
\omega _{b}$. We set $\protect\theta =0.8\protect\pi $, $g_{ma}/2\protect\pi %
=3.5$MHz and the other parameters we chose are the same as those in Fig.2.}
\end{figure}

It is also important to address the effect of noise on the studied
entanglement. This is investigated in Fig. 4 and Fig. 5, where the
entanglements $E_{\mathcal{N}\text{,}am}$ and $E_{\mathcal{N}\text{,}bm}$
are plotted versus the temperature $T$, respectively. Both entanglements are
sensitive to the temperature $T$ and the strong entanglements only exist at
cryogenic temperatures. And the entanglements increase with the increase of $%
\chi $. Here we choose $\chi $ that are less than the optimal value in Fig.
2.

\begin{figure}[tbp]
\centering\includegraphics[width=6cm]{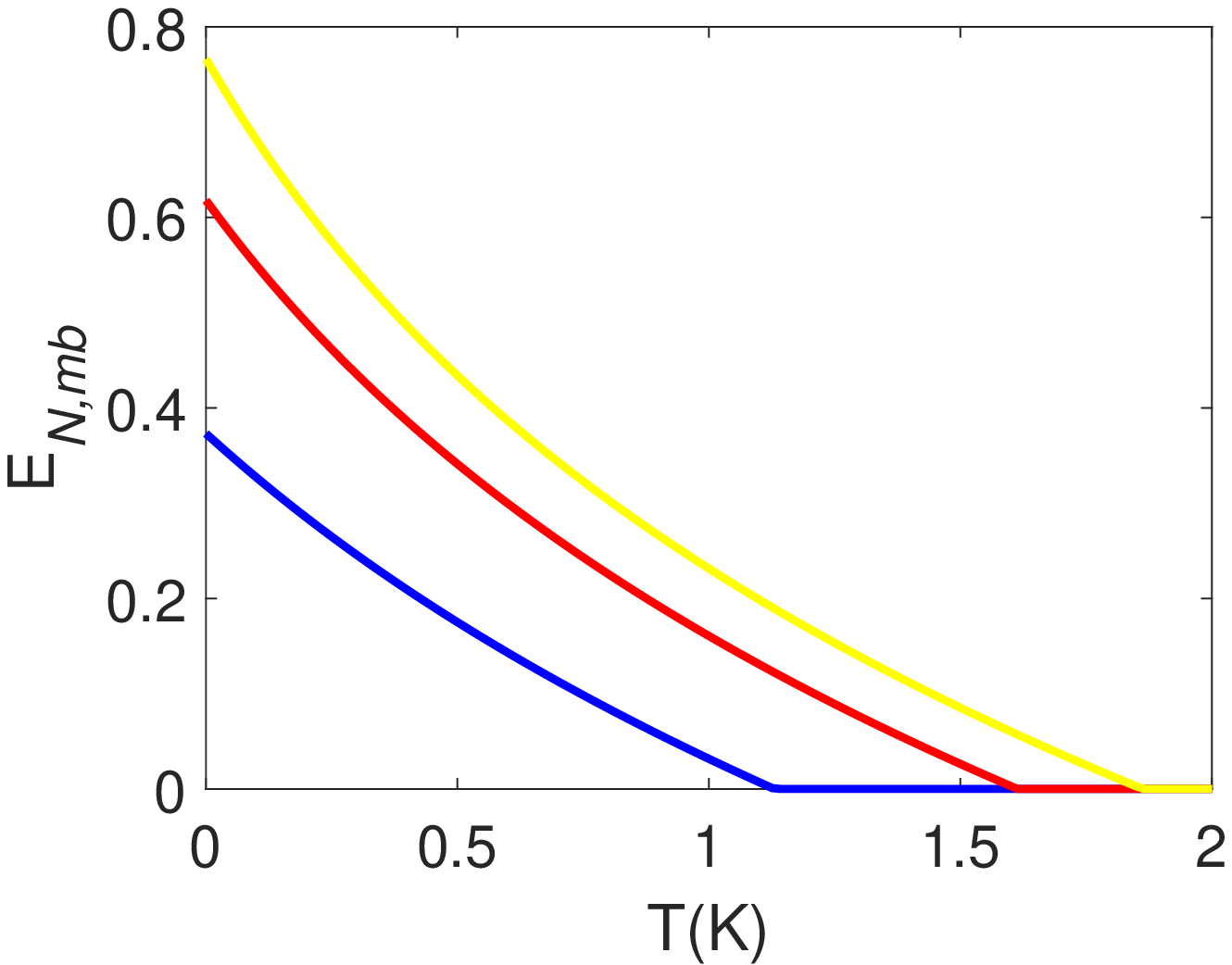} 
\caption{The entanglements $E_{\mathcal{N}\text{,}bm}$ as functions of the
temperature $T$. The yellow line denotes the squeezing parameter $\protect%
\chi =0.15\protect\omega _{b}$, the red line denotes $\protect\chi =0.1%
\protect\omega _{b}$ and the blue line denotes $\protect\chi =0.04\protect%
\omega _{b}$. We set $\protect\theta =0.8\protect\pi $, $g_{ma}/2\protect\pi %
=4.7$MHz and the other parameters we chose are the same as those in Fig.2.}
\end{figure}

Lastly, we discuss the implementation of the experimental in the CMM system.
Based on the development of the material science, we assume that an active
cavity mode can be constructed by the doping active metamaterials with
inherent enhancement in the cavity. This assumption is based on two existing
works:(1) the $\mathcal{PT}$-symmetric whispering-gallery microcavitie \cite%
{k18}. (2) the acoustic gain by nonlinear active acoustic metamaterials \cite%
{k47}. In addition, the bipartite entanglements can be measured by the
cavity field quadratures, which can be measured directly by the output of
the cavity, and the magnon state can be measured indirectly. By an auxiliary
optical cavity which is considered to couple the YIG sphere and the
quadratures of the mechanical mode can be obtained \cite{k7,k44}. Here, we
set $G>\kappa _{m}$ to ensure that unwanted magnon Kerr effect is ignored in
a strong driving field of magnon mode \cite{k8,k62}.

\section{Conclusions}

In summary, we have investigated the enhancement of the bipartite
entanglements in a CMM system via the squeezing of the magnon mode. The
squeezing leads to the enhancement of nonlinearity in the system, which
leads to the increase of entanglement, especially for the case of the low
temperature. And the enhancement is not a simple linear enhancement, there
is an optimal parameter range, which corresponds to the optimal enhancement
effect on the entanglement. Since a novel squeezing of the magneon mode is
considered, we mainly focus on entanglement related to magnon mode, such as
the photon-magnon and magnon-phonon entanglements. In addition, the
experimental implementation is also discussed. We expect that the proposed
scheme provides provides an alternative method to enhance entanglement in
the CMM systems and it has potential applications in the study of
macroscopic quantum state and quantum information network.

\section*{DISCLOSURES}

The authors declare no conflicts of interest.

\bibliographystyle{unsrt}
\bibliography{0528.bib}

\end{document}